\documentclass[fleqn]{article}  % twocolumn % affiliation des auteurs
\usepackage[utf8]{inputenc}
\usepackage{caption}
\usepackage[T1]{fontenc}
\usepackage{graphicx}
\usepackage{amsmath}
\usepackage{amssymb}
\usepackage{amsfonts}
\usepackage{esvect}
\usepackage{ifthen}
\usepackage{mathtools}
\usepackage{kpfonts}
\usepackage{array,multirow,makecell}
\setcellgapes{1pt}
\makegapedcells
\newcolumntype{R}[1]{>{\raggedleft\arraybackslash }b{#1}}
\newcolumntype{L}[1]{>{\raggedright\arraybackslash }b{#1}}
\newcolumntype{C}[1]{>{\centering\arraybackslash }b{#1}}
\title{Cherenkov Telescope Arrays for High-Energy Neutrino Detection in Mountain Ranges}
\author{Tariq BITAM*, Bouzid BOUSSAHA*,Mossaab MESSAMAH*}
\date{}
\begin{document}
\maketitle{}
%\begin{multicols}{2}
*Department of sciences matter , University of Algiers, Benyoucef Benkhedda, 2 Rue Didouche Mourad, Algeria.\\
\begin{abstract}
The present study investigates the feasibility of employing Cherenkov Telescope Array (CTA) technology for the detection of ultra-high-energy (UHE) neutrino-tau particles. By observing the Cherenkov light produced by charged particles resulting from neutrino-tau interactions within mountainous regions, the research seeks to assess the viability of detecting neutrinos within the  $10^{16}-10^{18}eV$ GeV energy range. The present paper details the methodology used to calculate the effective detection area and the interaction probabilities of neutrino-tau within the mountains, as well as the flux of Cherenkov photons generated by tau leptons. Furthermore, the present paper addresses the selection of optimal parameters for the telescopes, including mountain dimensions, telescope height, the distance between telescopes, and the distance from the telescopes to the mountain. Strategic placement of the telescopes and the identification of mountain ranges in Algeria that are most suitable for detecting these elusive particles are also discussed. The findings and recommendations for the optimal distribution of telescopes across the selected mountain range are presented, aiming to enhance the detection capabilities for high-energy neutrino-tau particles.
\end{abstract}
\section*{Introduction}
The detection of ultra-high-energy (UHE) neutrinos has significantly advanced the field of astronomy, largely due to the contributions of the IceCube Collaboration [1]. Unlike photons or charged cosmic rays, UHE neutrinos do not interact with other particles, enabling astronomers to study the cosmic microwave background (CMB) radiation within the   $10^{16}-10^{18}eV$, energy range, and to investigate the inner solar reactions of the Sun [2]. Neutrinos are the most abundant particles in the universe and play a crucial role in its evolution. It is hypothesized that a difference in the mixing behaviors of neutrino types compared to antineutrino types may explain why matter prevails over antimatter in the universe, thereby explaining our existence. 

The two primary challenges in detecting ultra-high-energy (UHE) neutrinos are their minuscule cross-section and extremely low flux. These challenges are addressed by instrumenting large detection volumes in a cost-effective manner. Currently, efforts to detect UHE neutrinos are divided into two main approaches. The first approach involves using large volumes of ice or water, where a neutrino interacts and produces a particle shower. This method is employed by experiments such as IceCube and ANTARES/KM3Net [3, 4]. Notably, IceCube has demonstrated sensitivity to UHE neutrinos, and its published limits are the most stringent up to approximately  $10^{11}GeV$  [5].The second approach is to use ice as a detection medium and look for the radio signature of a particle shower, either by deploying radio antennas in ice (ARA[6, 7]., ARIANNA[8,9].) or by observing from a balloon (ANITA)[10]. For reviews about the detection of air showers with radio and a more complete list of experiments that use or plan to use radio for the detec- tion of UHE neutrinos, see [11, 12].
The second approach involves the so-called earth-skimming technique [13]. In this method, an ultra-high-energy (UHE) tau neutrino interacts with matter inside the Earth after traveling a distance of tens to hundreds of kilometers. This interaction produces a tau particle that propagates through the Earth and may emerge from the surface if it does not decay beforehand. Upon emergence, the tau particle decays in the atmosphere, initiating an air shower that can be detected [14-17]. Although neutrinos interact very weakly with matter, they can produce observable effects if they encounter a substantial amount of matter, such as mountain ranges. In such scenarios, neutrinos can generate tau particles that create a cascade of high-energy charged particles, which emit Cherenkov radiation in the atmosphere. This study will investigate the feasibility of using an apparatus such as the Cherenkov Telescope Array (CTA) to detect earth-skimming UHE neutrino-tau particles within the energy range of  $10^{7}-10^{8}$ GeV by tracking the Cherenkov light flux generated by the charged particles resulting from neutrino-tau conversion within mountain ranges [18]. Additionally, this research will aim to determine the optimal parameters for the collectors, their locations, and the properties of the mountain ranges that provide the best opportunities for capturing these high-energy neutrino-tau particles.
The present paper investigates the detection of tau neutrinos in a mountainous region using Cherenkov telescopes. Section II introduces the concept of the effective detection area. Subsequently, in Section III, the probability of interaction and the effective detection area of tau neutrinos within the mountain is computed. Section IV reviews the calculation of the flux of Cherenkov photons produced by tau leptons, considering various geographical and telescopic parameters. It also calculates the effective area for capturing tau leptons. Section V identifies a suitable mountain range in Algeria and proposes an optimal distribution of telescopes. In conclusion, Section VI discusses the results and conclusions.

\section{Effective detection area}
Let us designate by $A_{eff}$ the effective   area of the Imaging Atmospheric Cherenkov Telescope (IACT), in consideration, for the detection of a shower induced by a tau with energy  $E_{\tau}$exiting the Earth with a downward direction denoted by ${\Omega}$,and emerging from coordinates  (x,y).). Hence, the total number of detected events can be expressed as:

\begin{equation}
$$
\frac{dN\tau}{dt}=\int dE_{\tau} d{\Omega} \frac{d{\Phi}_{\tau}(E_{\tau},{\Omega})}{dE_{\tau}d{\Omega}}A^{\tau}_{eff}(E_{\tau},{\Omega})=\int dE_{\nu} d{\Omega} \frac{d{\Phi}_{\nu}(E_{\nu},{\Omega})}{dE_{\nu}d{\Omega}}A^{\nu}_{eff}(E_{\nu},{\Omega})
$$
\end{equation}

The quantity $A_{eff}^\nu\left(E_\nu,\Omega\right)$ h represents  the $\nu_\tau$ effective area of the apparatus and it is defined as
\begin{equation}
$$
A_{eff}^\nu\left(E_\nu,\Omega\right)=\int{\ dE_\tau{\ A}_{eff}^\tau\left(E_\tau,\Omega\right)\ \ K}(E_\nu,E_\tau,\Omega)
$$
\end{equation}
Where  $ K(E_\nu,E_\tau,\Omega)$the integral kernel, must be calculated considering the processes that generate the downward tau flux.

\section{Effective detection area of neutrino-tau}
A computational program was developed to estimate the effective area for capturing tau neutrinos utilizing the Monte Carlo method. This approach entails the random generation of $N_{gen}$ neutrinos-tau in the vicinity of a mountain range characterized by a length  L, width P, and height H. The tau neutrinos enter the mountain with an incidence angle ranging between 85 and 95 degrees, and a transverse angle spanning from 0 to 360 degrees. Within the mountain, these neutrinos traverse a distance $D_\nu $ before interacting to produce a tau lepton. The distance $D_\nu $ is determined by the following equation[19]:
\begin{equation}
$$
D_\nu=-\frac{\ln(R)}{\left(\sigma_{CC}+\sigma_{CN}\right)\rho N_A}
$$
\end{equation}
Where R is a random number between 0 and 1; $\sigma_{CC}$ and  $\sigma_{CN}$  represent the cross sections of charged current and neutral current interactions, respectively, as given by formulas  [20,21].
\begin{equation}
$$
\sigma_{CC}(E)=2.4\times\sigma_{CN}(E)=6.04\times10^{-36}cm^{2}(\frac{E}{GeV})^{0.358}
$$
\end{equation}

If $D_\nu$ is sufficiently large such that the neutrinos do not interact within the mountain range, they are considered lost. Conversely, if $D_\nu$ is smaller than the dimensions of the mountain, the tau neutrinos interact within the mountain. In this scenario, the tau lepton produced by the neutrino-tau is tracked. Should this lepton interact within the mountain, the tau neutrino is also deemed lost. The survival probability of the tau lepton is then determined by the following relationship [22].

\begin{equation}
$$
P_{surv}(E_{\tau},E_{\tau}^{i})=\exp\Bigg[\frac{m_{\tau}\beta_{1}}{c\tau\rho\beta_{0}^{2}}\bigg(\frac{1}{E_{\tau}}(1+\ln(\frac{E_{\tau}}{E_{0}}))-(\frac{1}{E_{\tau}^{i}}(1+\ln(\frac{E_{\tau}^{i}}{E_{0}}))\bigg)\Bigg]\exp\Bigg[-\frac{m_{\tau}}{c\tau\rho\beta_{0}}(\frac{1}{E_{\tau}}-\frac{1}{E_{\tau}^{i}})\Bigg]
$$
\end{equation}
\begin{equation}
$$
E_{\tau}=\exp\Bigg[-\frac{\beta_{0}}{\beta_{1}}\bigg(1-e^{-\beta_{1}\rho Z}\bigg)+\ln\bigg(\frac{E_{\tau}^{i}}{E_{0}}\bigg)e^{-\beta_{1}\rho Z}\Bigg]E_{0}            
$$
\end{equation}
where;\\
\begin{equation}
$$
\left\lbrace
	\begin{aligned}
\beta &= \beta_{0}+\beta_{1}ln\bigg(\frac{E}{E_{0}}\bigg) \\
\beta_{0} &=1.2\times10^{-6}cm^{2}g^{-1} \\
\beta_{1} &=0.16\times10^{-6}cm^{2}g^{-1}
	\end{aligned}
	\right.
$$
\end{equation}
In these expressions, $Z$ denotes the distance traveled by the $\tau$ following its production,$E_{\tau}^{i}$ represents  the initial $\tau$ energy , precisely at moment of its production  with $E_{\tau}^{i}=0.8E_{\nu}$ and $E_{0}=10^{10}GeV$. 

If the tau lepton survives and exits the mountain range, subsequently interacting in the atmosphere within a distance  x less than 50 kmkm in the vicinity of the mountain range and at an altitude below 10 km, the neutrino is considered capturable [23]. However, if the lepton interacts outside these specified dimensions, the tau neutrino is considered lost and cannot be captured. The total number of capturable tau neutrinos is denoted by  $N_k$. In order to calculate the effective area for tau neutrinos, we employ the following relationship:

\begin{equation}
$$
A\left(E_{\nu_\tau}\right)=N_{gen}^{-1}\sum_{i=1}^{N_k}{P_i\left(E_{\nu_\tau},E_\tau,\theta\right)}T_{eff,i}(E_\tau,x,y,h,\theta)A_i(\theta)\Delta\Omega
$$
\end{equation}

where $N_{gen}$ represents the number of generated neutrino-tau events. $N_k$ is the number of  $\tau$  leptons with energies $E_\tau$  larger than the threshold energy $E_{th}$ = 1 PeV. 
$P_i\left(E_{\nu_\tau},E_\tau,\theta\right)$ represents the probability that a tau neutrino with energy  $E_{\nu_\tau}$ and zenith angle $\theta$   produces a lepton tau with energy $E_\tau$.
$A_i(\theta )$ denotes  the physical cross-section of the interaction volume experienced  by the neutrino-tau and $\Delta\Omega$  represents the solid angle.

$T_{eff,i}(E_\tau,x,y,h,\theta)$ is the trigger efficiency for tau-lepton induced showers with the decay vertex positioned at (x,y) and height h above the ground
to calculate the interaction probability of a  neutrino-tau with energy $E_{\nu_\tau}$ and the emergence probability of a tau with energy $E_\tau$ the formula is used [24]
\begin{equation}
$$
P_i\left(E_{\nu_\tau},E_\tau,\theta\right)=\frac{1}{\beta\rho E_\tau}P_{\nu_\tau} P_\tau
$$
\end{equation}
where the neutrino-tau interaction probability is:
\begin{equation}
$$
P_{\nu_\tau}=\sigma_{cc}\rho N_A\exp[-D_\nu\left(\sigma_{CC}+\sigma_{CN}\right)\rho N_A)]
$$
\end{equation}
and the probability for the tau to emerge from the ground is:
\begin{equation}
$$
P_\tau=\exp[-\frac{m_\tau}{\left(\tau_\tau c\beta\rho E_\tau\right)}(1-e^{-\beta\rho D_\tau})]
$$
\end{equation}
where $\rho$ = 2,65 g/cm3 is the density of rock,$N_A$ is the Avogadro constant, $\tau_\tau$ is the lifetime of the tau,$ m_\tau$ is its mass, and c is the speed of light.

$D_\tau$  denotes the distance the tau lepton must travel through the mountain before it emerges, and the energy upon emergence from the ground is assumed to be $E_\tau$.
In the simulation program we developed, one million high-energy neutrinos, with energies ranging between  $10^{16} and 10^{18}$ eV  are generated. Their entry angles are varied between 85 and 90 degrees around mountain ranges of different dimensions. Ultimately, we obtain the interaction coordinates of the tau leptons generated by the neutrinos after they exit the mountain mass, within the traceable limits previously mentioned. Figure 1 provides an example of these results.

\begin{figure}[h]
\centering	
\includegraphics[scale=0.7]{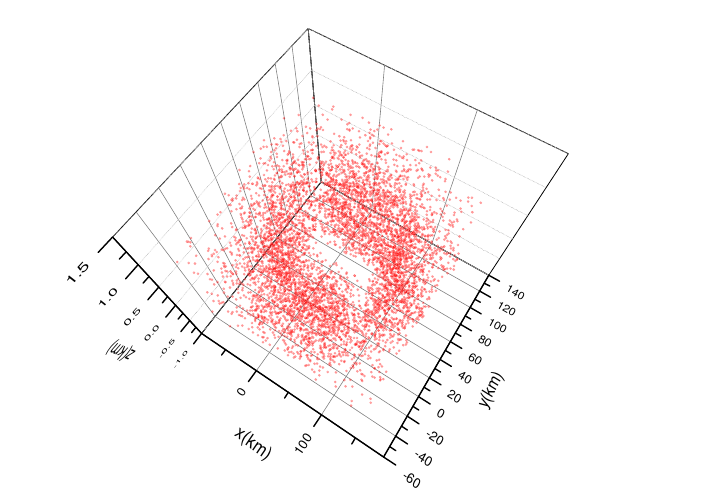}
\caption{the coordinates (x,y,z)of the Tau leptons emerging and decaying within the  interaction volume}
\end{figure}

The program also facilitates the calculation of the effective area for detecting neutrinos as a function of the various dimensions of the mountainous masses. and figure 2 illustrates the variation in the effective area with respect to changes in the dimensions of the mountainous mass. 
\begin{figure}[h]
\centering	
\includegraphics[width=0.70\linewidth]{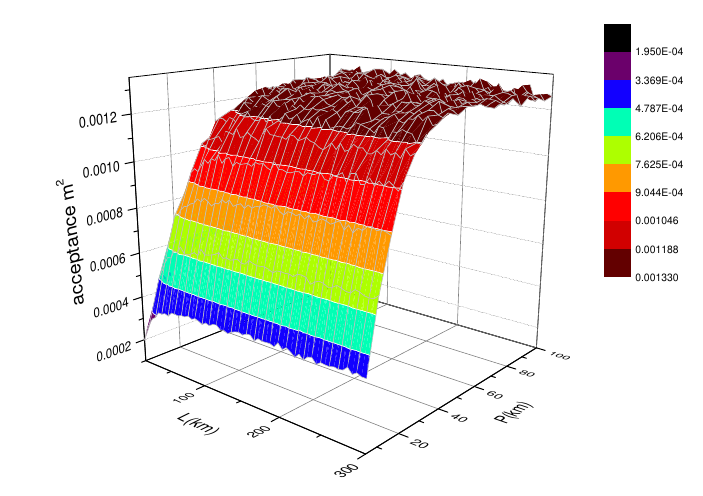}
\caption{Variation in the acceptance for tau neutrinos as a function of the dimensions of the mountain range, presented in a 3D format with a height of 2 km. }
\end{figure}
 Through the analysis of figure 3 we determined that the red zone, indicating maximum acceptance, begins at a minimum of (L*P$\geq$ 40*40 $km^{2}$  )
\begin{figure}[h]
\centering	
\includegraphics[width=0.70\linewidth]{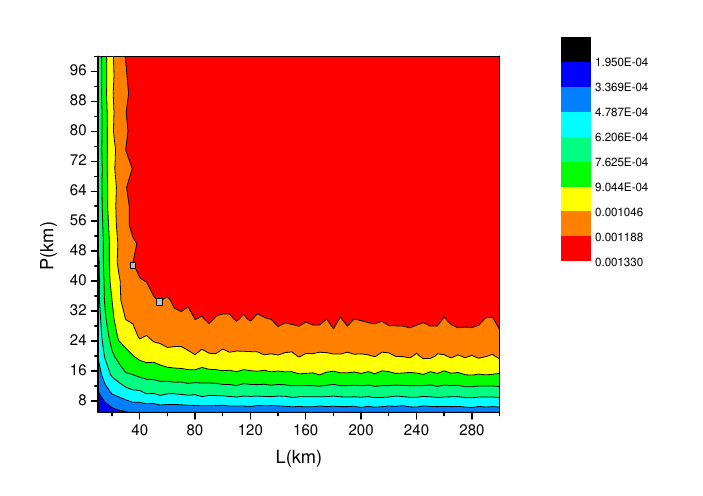}
\caption{The acceptance of tau neutrino detection as a function of the dimensions of the mountain range, depicted in contour format.}
\end{figure}

\section{Effective detection area of  lepton tau}

For the purpose of calculating the effective area for capturing the tau lepton produced by a high-energy neutrino, we utilize the CORSIKA program [25]. This program employs the Monte Carlo algorithm to track the path of the tau lepton and the charged particles and photons generated from it. Our study focuses on tracking Cherenkov photons generated by the tau lepton or by high-energy particles produced from the tau lepton.
 In the simulation, we input the position and direction of the tau lepton in the atmosphere, assuming it has an energy  $E_\tau$ and an almost vertical entry angle (CORSIKA considers a maximum vertical angle of 88 degrees), at a height  H above sea level at the coordinates  (x = 0  , y = 0) The lepton's trajectory is assumed to lie in the z-x plane , directed towards the positive the  x axis. We consider the physical phenomena to be symmetric with respect to the   z-x plane, , and we track the Cherenkov photons resulting from the tau lepton's path in the atmosphere based on its initial energy. Telescopes are arranged in an array on various y-z planes intersecting the  x axis at different distances (Figure 4).

  This network of telescopes enables us to obtain simulation results determining the flux of Cherenkov photons relative to the surface at each telescope. Using interpolation, Cherenkov photons can be tracked at every point in the atmosphere.
We determine the effective area of the telescope by identifying the minimum number of Cherenkov photons that the telescope can capture. We then search the simulation results for the optimal level that achieves the largest possible area for capturing Cherenkov photons, which we consider the effective area of the telescope.
In order to determine the optimal values for the parameters of these col- lectors and the best places for their positioning, we search within the previous level, which was formed by a matrix of telescopes, for the best height for tele- scopes that achieve the highest field for capturing Cherenkov photons, and we consider it the optimal height above sea level for the positioning of telescopes. In the simulation, the CORSIKA program provides the number of Cherenkov photons relative to the depth from the starting point of the tau lepton. This data enables us to determine the maximum extent of this phenomenon,which in turn helps us to establish the minimum distance along the x axis in which telescopes can be placed.

   \begin{figure}[h]
\centering	
\includegraphics[width=0.85\linewidth]{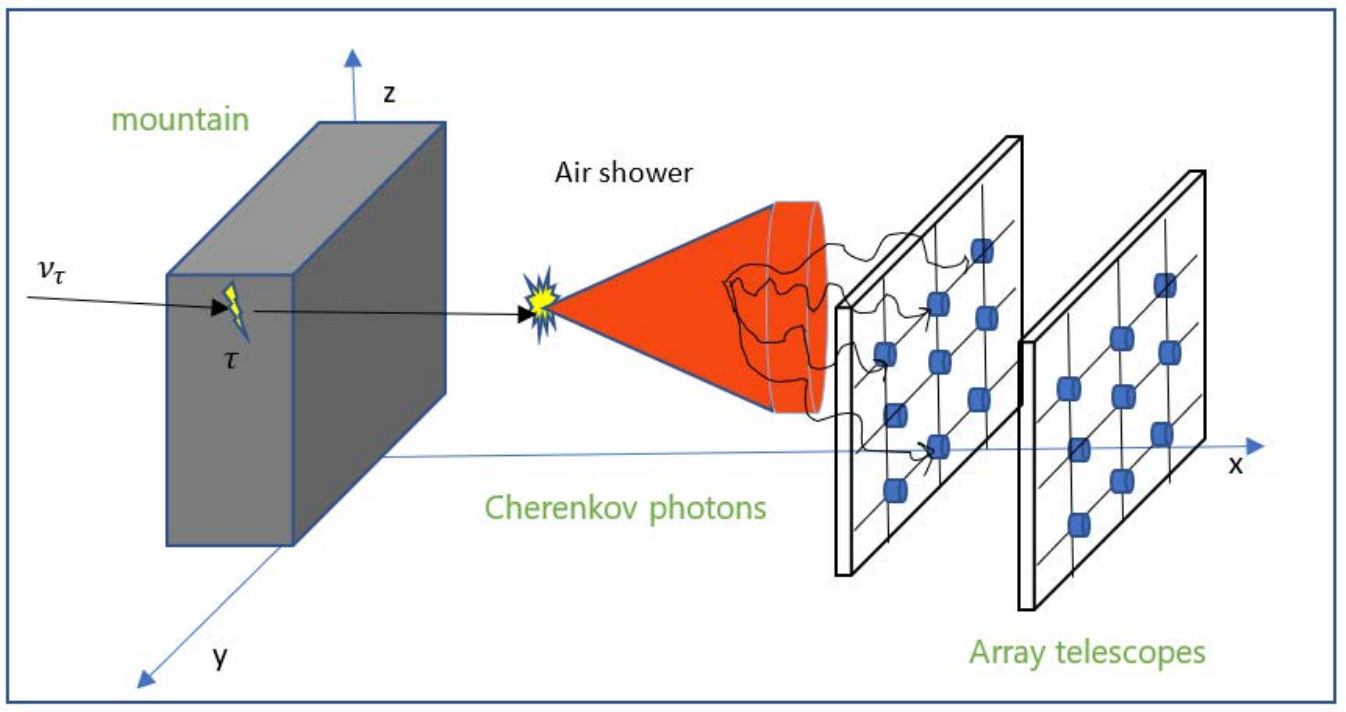}
\caption{Illustration depicting the tracking of tau leptons utilized in our CORSIKA simulation.}
\end{figure}
   
\begin{figure}[h]
	\begin{minipage}[b]{0.60\linewidth}
		\centering \includegraphics[scale=0.6]{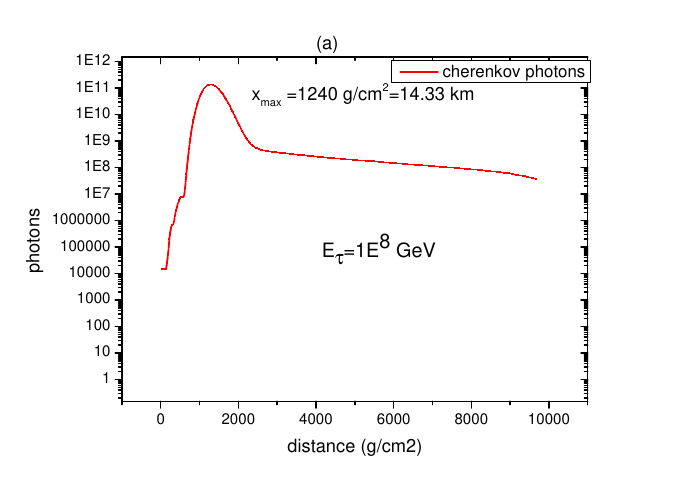}
		%\caption{\it Légende 1}
	\end{minipage}\hfill
	\begin{minipage}[b]{0.60\linewidth}	
		\centering \includegraphics[scale=0.6]{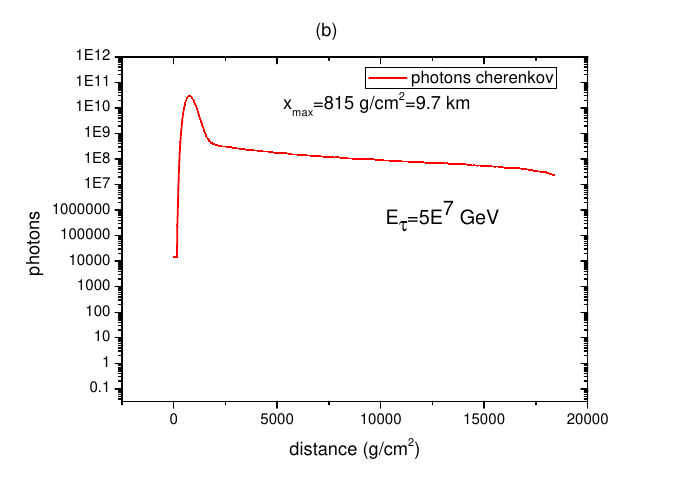}
		%\caption{Légende 2}
	\end{minipage}
	\begin{minipage}[b]{0.60\linewidth}
		\centering \includegraphics[scale=0.6]{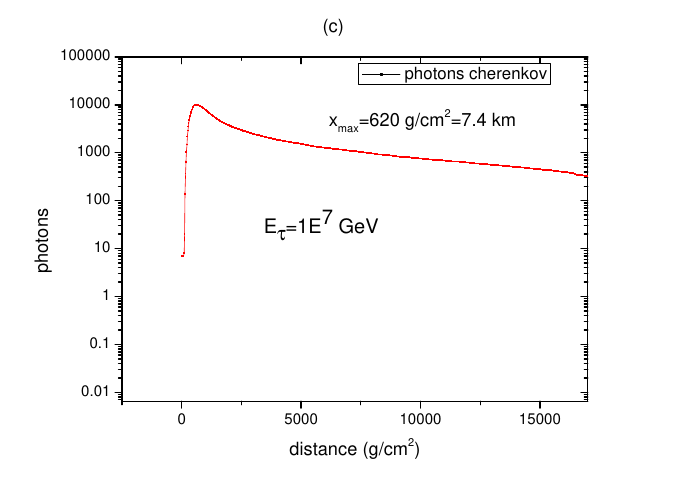}
		%\caption{Légende 3}
	\end{minipage}\hfill
	\caption{The longitudinal profile for Cherenkov photons from showers initiated by tau leptons as a function of atmospheric depth. Subfigures depict profiles for different tau lepton energies ((a):$E_\tau= 10^{8}$ GeV, (b): $E_\tau=5 10^{7}$ GeV, (c): $E_\tau=10^{7}$ GeV )}
\end{figure}
\newpage

In Figure 5 the curves (a), (b), and (c) , represent the number of Cherenkov photons resulting from the interaction of high-energy tau particles and their products with the atmosphere after their exit from the mountains, plotted as a function of the distance of their penetration. These curves correspond to tau lepton energies of  $10^{8},5 10^{7},10^{7}$ GeV , respectively.
Observing curve (a), we find that the number of Cherenkov photons reaches its maximum at a depth of  1240 g/$cm^{2}$², corresponding to a distance of 14.33 km. In curve (b), the peak occurs at 815 g/$cm^{2}$ ², equivalent to a distance of 9.77 km. For curve (c), the peak is observed at a depth of  620 g/$cm^{2}$ ², which corresponds to 7.4 km. The summary of peak points for each energy along the x-axis is as follows:

\begin{center}
\begin{tabular}{|L{3.5cm}||L{1.75cm}|L{1.75cm}|L{1.75cm}|}
\hline energy $ E_\tau=$ & $10^{7}$ GeV & $5 10^{7}$ GeV &  $10^{8}$ GeV \\
\hline  Slant distance ($x_{max}$) & 7.4 km & 9.77 km & 14.33 km  \\
\hline 
\end{tabular}
\captionof{table}{the peak points of Cherenkov phenomena for each  tau energy at x  axis}
\label{tab1}
\end{center}
In this section, our objective is to explore the effective area for detecting tau particles concerning their altitude above sea level upon emergence from the mountain range, assuming a perpendicular angle of incidence, and considering their initial energy. We choose two representative altitudes for the tau particles based on specific criteria: 
1.	Low mountain ranges of 2 km: These include regions such as those found in Africa, Europe, and Australia.
2.	Medium mountain ranges of 4 km: Such as those prevalent in Asia.
To compute the flux of Cherenkov photons, we utilize the CORSIKA program. We employ an array of spherical telescopes, each with an area of 10 m², positioned at various distances from the peak point of the Cherenkov photons, depending on the initial energy of the tau leptons. The results of these computations are presented in Figures 6 and 7.

\begin{figure}[h]
	\begin{minipage}[b]{0.60\linewidth}
		\centering \includegraphics[scale=0.6]{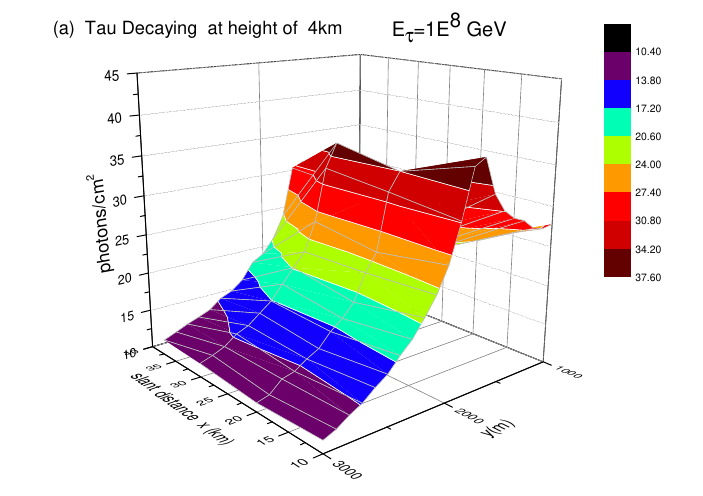}
		%\caption{\it Légende 1}
	\end{minipage}\hfill
	\begin{minipage}[b]{0.60\linewidth}	
		\centering \includegraphics[scale=0.6]{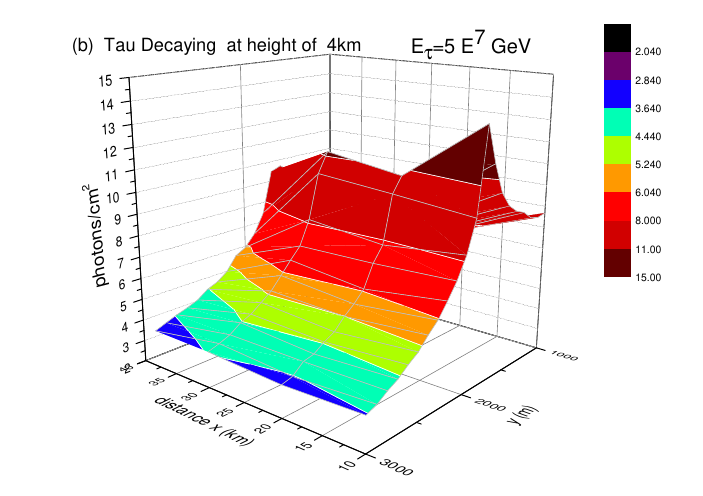}
		%\caption{Légende 2}
	\end{minipage}
	\begin{minipage}[b]{0.60\linewidth}
		\centering \includegraphics[scale=0.6]{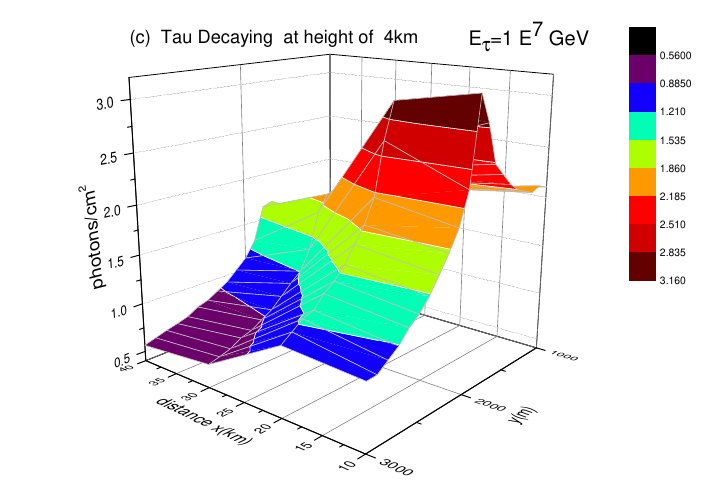}
		%\caption{Légende 3}
	\end{minipage}\hfill
	\caption{:  The distribution of Cherenkov photons flow depicted in terms of the x and y coordinates, representing the horizontal and vertical distances from the initial position of the tau leptons, for various energy values decaying at an emerged height of 4 km }
\end{figure}
\newpage
   
\begin{figure}[h]
	\begin{minipage}[b]{0.60\linewidth}
		\centering \includegraphics[scale=0.6]{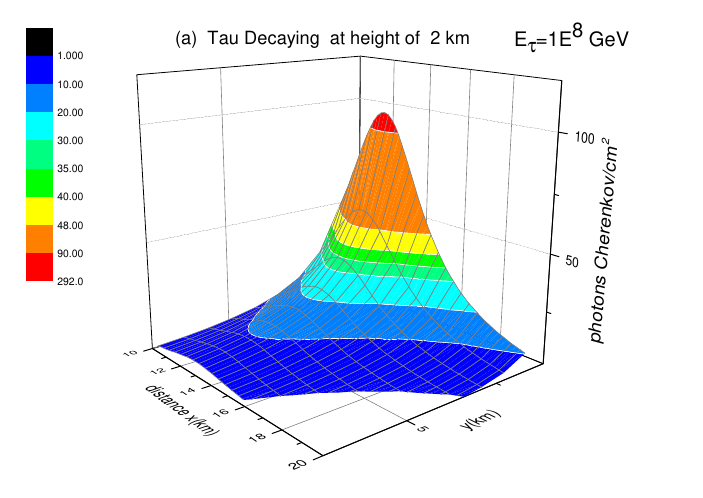}
		%\caption{\it Légende 1}
	\end{minipage}\hfill
	\begin{minipage}[b]{0.60\linewidth}	
		\centering \includegraphics[scale=0.6]{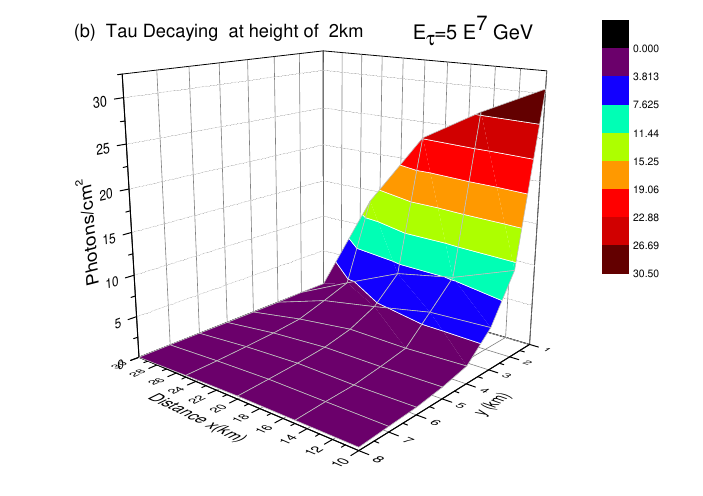}
		%\caption{Légende 2}
	\end{minipage}
	\begin{minipage}[b]{0.60\linewidth}
		\centering \includegraphics[scale=0.6]{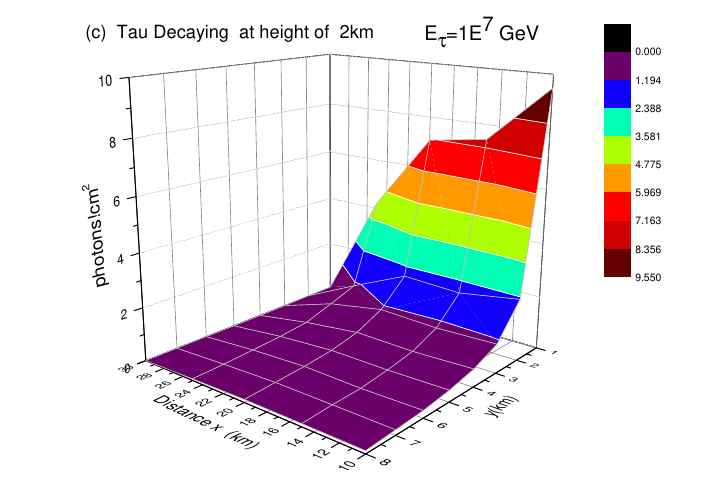}
		%\caption{Légende 3}
	\end{minipage}\hfill
	\caption{The distribution of the Cherenkov photon flux as a function of the x and y coordinates, which represent the horizontal and vertical distances from the initial position of the tau leptons, is analyzed for various energy levels, decaying at an emerged height of 2 km   }
\end{figure}

At the positions of the peak points (Table 1) of the Cherenkov photons, the height of the telescopes along the oz axis is adjusted to search for the peak value and the minimum value that the telescope can detect as a function of the energy and initial position of the tau lepton. The results are depicted in Figure 8.
\newpage
\begin{figure}[h]
	\begin{minipage}[b]{0.60\linewidth}
		\centering \includegraphics[scale=0.6]{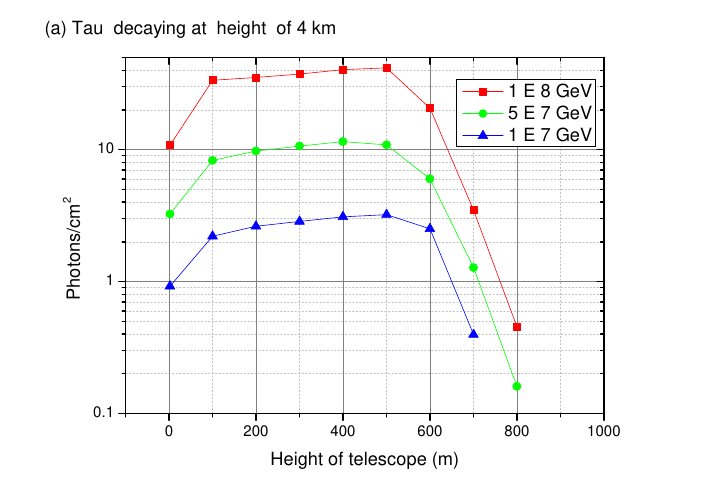}
		%\caption{\it Légende 1}
	\end{minipage}\hfill
	\begin{minipage}[b]{0.60\linewidth}	
		\centering \includegraphics[scale=0.6]{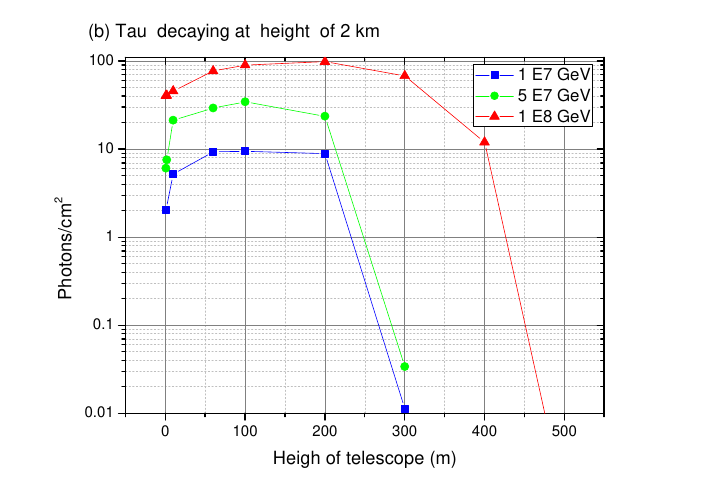}
		%\caption{Légende 2}
	\end{minipage}\hfill
	\caption{ Distribution of the Cherenkov photon flux for different energy values $E_\tau$ and two values of tau decaying height as a function of the telescope height: (a) Tau decaying at a height of 4 km, and (b) Tau decaying at a height of 2 km.)}
\end{figure}
Peak points for each energy are found at a height of 200 m for tau leptons decaying at a height of 2 km, and at a height of 500 m for tau leptons decaying at a height of 4 km. The minimum value that the telescope can capture is set to 1 photon/$cm^{2}$ ,The threshold height allowing the telescope to capture Cherenkov photons is summarized in Table 2.
\newpage
\begin{center}
\begin{tabular}{|R{3cm}||C{1.5cm}|L{1.5cm}|L{1.5cm}|}
\hline energy & $E_\tau=10^{7}$ GeV & $ E_\tau=5*10^{7}$GeV &  $E_\tau=10^{8}$ GeV \\
\hline  The threshold height (tau decaying at 2 km)  & 240 m & 250 m & 420 m \\
\hline  The threshold height (tau decaying at 4 km)  & 650 m & 700 m & 760 m \\
\hline 
\end{tabular}
\captionof{table}{The threshold height that allows telescope to capture Cherenkov photons for each  tau energy at  two height of tau decaying     }
\label{table3}
\end{center}	

Fixing the x and z coordinates of the telescopes (representing the peak coordinates from Table 1 and Table 2), the position of the telescope along the (oy) axis axis is adjusted to find the coordinate Y, corresponding to the lower limit value that the telescope can capture. Considering the symmetry of the Cherenkov photon distribution with respect to the (oxz) plane, twice the distance  $Y_{\th}$ is deemed the most suitable distance between two consecutive telescopes on the (oy) axis to observe neutrinos. The results are presented in Figure 9.
\begin{figure}[h]
	\begin{minipage}[b]{0.60\linewidth}
		\centering \includegraphics[scale=0.6]{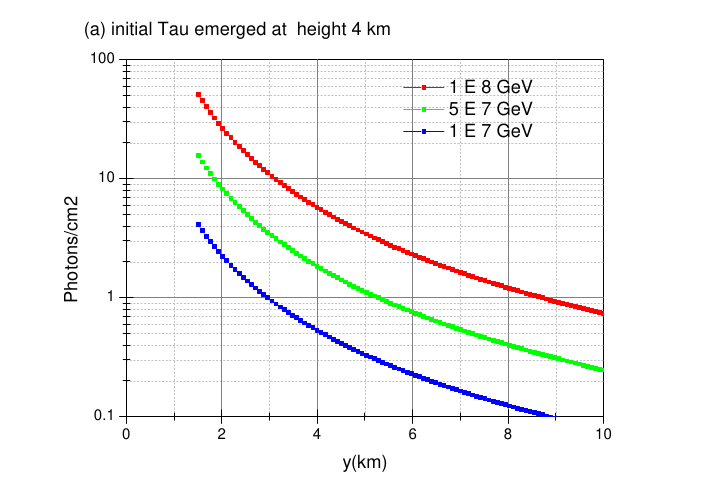}
		%\caption{\it Légende 1}
	\end{minipage}\hfill
	\begin{minipage}[b]{0.60\linewidth}	
		\centering \includegraphics[scale=0.6]{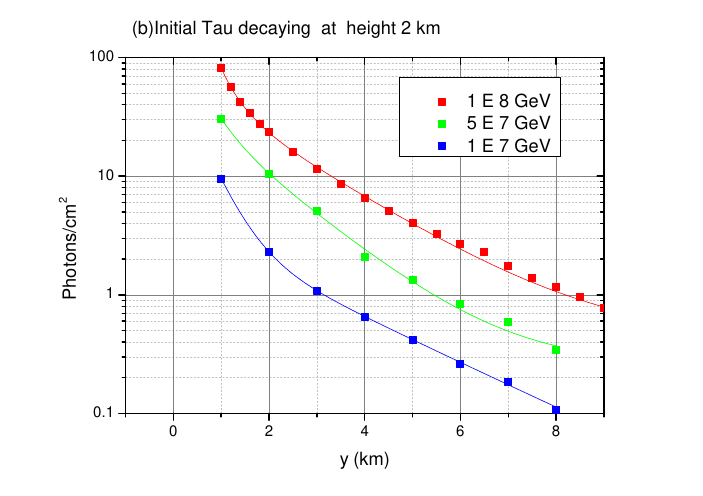}
		%\caption{Légende 2}
	\end{minipage}\hfill
	\caption{The distribution of Cherenkov photon flux along the y-axis is analyzed for different values of energy $E_\tau$and for two distinct tau decay heights: (a) tau decaying at a height of 4 km, and (b) tau decaying at a height of 2 km.}
\end{figure}
\newpage

By analyzing the results presented in the various figures, we summarize the optimal parameters for the placement of the telescopes based on the energy and initial position of the tau leptons, as detailed in Table 3 below:
\begin{center}
\begin{tabular}{|C{2.5 cm}||C{2.5 cm}|C{2.7 cm}|}
\hline \multirow{2}*{} & \multicolumn{2}{l|}{Distance (telescope - mountain) }\\
\cline{2-3}  &  tau  decaying height at 2 km  & tau  decaying height at 4 km  \\ 
\hline $E_\tau=10^8GeV$  & 15 km & 15 km  \\
\hline $E_\tau=5*10^7GeV$  & 12 km & 12 km  \\
\hline $E_\tau=10^7GeV$  & 12 km & 12 km  \\
\hline
\hline \multirow{2}*{} & \multicolumn{2}{l|}{Height of telescope  }\\
\cline{2-3}   &  tau  decaying height at 2 km  &  tau  decaying height at 4 km  \\  		
\hline $E_\tau=10^8GeV$  & <420 m & <760 m \\
\hline $E_\tau=5*10^7GeV$  &< 250 m & <700 m \\
\hline $E_\tau=10^7GeV$  & < 240 m	 & < 650 m  \\ 
\hline
\hline \multirow{2}*{} & \multicolumn{2}{l|}{Distance between two  telescopes }\\
\cline{2-3}  &  tau  decaying height at 2 km  &  tau  decaying height at 4 km  \\ 
\hline $E_\tau=10^8GeV$  & 16 km & 17.5 km  \\
\hline $E_\tau=5*10^7GeV$  & 10.5 km & 11 km  \\
\hline $E_\tau=10^7GeV$  & 6 km & 6 km  \\ 
\hline
\end{tabular}
\captionof{table}{the optimal parameters for the location of the telescopes according  for different values  of energy  $E_\tau$  and two values of tau  decaying height position (2 and 4 km)   }
\label{table3}
\end{center}
These parameters enable the optimal capture and tracking of tau neutrinos. Additionally, by determining the minimum capture limits, the effective area of the telescope, as depicted in Figure 10 can be calculated.

\begin{figure}[h]
	\centering
	   \includegraphics[width=0.70\linewidth]{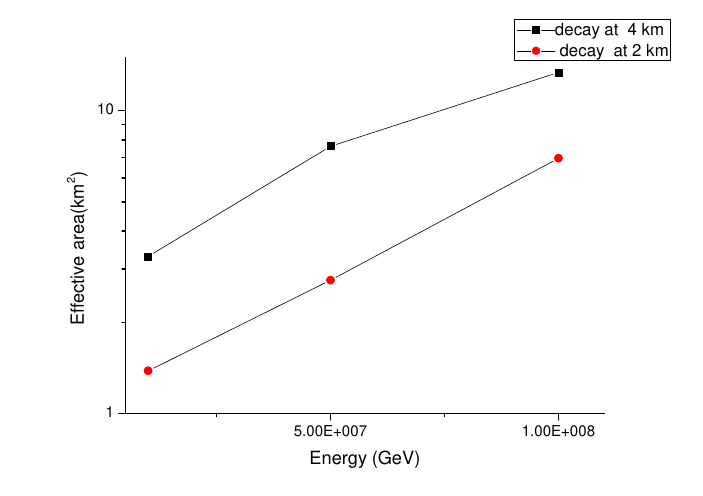}
	\caption{The effective area of detection of lepton tau according to the energy initial and the height of premier interaction  in atmosphere of the lepton tau }
\end{figure}

 Based on the results shown in Figure 3 and the data in Table 3, we identified the optimal parameters for tracking tau neutrinos with high energies  $E_\tau\geq$ $ 10^{8} $GeV The optimal parameters are summarized in Table 4 : 
\begin{center}
\begin{tabular}{|R{3cm}||C{2.5cm}|L{2.5cm}|L{2.5cm}|}
\hline Mountainous dimensions  & Telescopes height  & Distance between two telescopes& Distance (telescope - mountain) \\
\hline $ \geq$ 40 * 40$ km^{2}$  &  $ \leq$300-400 m & 12 -17 km & 10 -18 km \\
\hline 
\end{tabular}
\captionof{table}{optimal parameters  for traking the neutrinos tau of height energies   }
\label{table4}
\end{center}	
\section{Proposition of suitable mountain range in Algeria for traking neutrinos tau }
Given our presence in Algeria and the abundance of mountain ranges in the country, we conducted a search for a location meeting the criteria outlined in Table 4. Specifically, we sought mountain ranges with dimensions exceeding  40 * 40 $km^{2}$and heights surpassing 1500 m. Additionally, we considered the presence of depressions within 15 km of these mountain ranges, with depths less than 400 m. Our investigation led us to identify the Aures mountain range as meeting these requirements. Upon analyzing the topographical map of the Aures region, we propose distributing telescopes in the southern area of this mountain range. This strategic positioning aims to optimize conditions for tracking neutrinos, ensuring comprehensive capture of neutron traces resulting from interactions with the mountain ranges. We determined that deploying seven telescopes in the designated area (refer to Figure 11 [26-27]) would enable effective tracking of neutrons. Table 5 provides detailed information on these designated areas.
\begin{center}
\begin{tabular}{|R{1cm}||C{2.5cm}|L{3.5cm}|L{2.5cm}|}
\hline N & local  & Coordinates & Altitude \\
\hline 1 &	 Lutaya, Biskra &	35°01'39"N5°35'15"E	& 250 m\\
\hline 2 &	Aqaba Boumouche,Biskra &	34°54'35"N5°41'41"E	& 170 m\\
\hline 3 &	Bouhanan, Biskra &	34°55'26"N5°58'53"E	& 270 m\\
\hline 4 &	   Sidi Okba,Biskra	& 34°47'14"N5°56'55"E &	65 m\\
\hline 5 &	 Ain ennaga, Biskra	& 34°42'26"N6°06'33"E &	20 m\\
\hline 6 &	 Jalal, Khenchela & 	34°47'52"N6°53'06"E &230 m\\
\hline 7 & Babar Khenchela	& 34°34'46"N7°16'02"E &	150 m\\
\hline 
\end{tabular}
\captionof{table}{Proposal for the optimal positioning of telescopes around the Aures mountain range for tracking neutrinos in Algeria.}
\label{table5}
\end{center}
\begin{figure}[h]
	\centering
	   \includegraphics[width=0.70\linewidth]{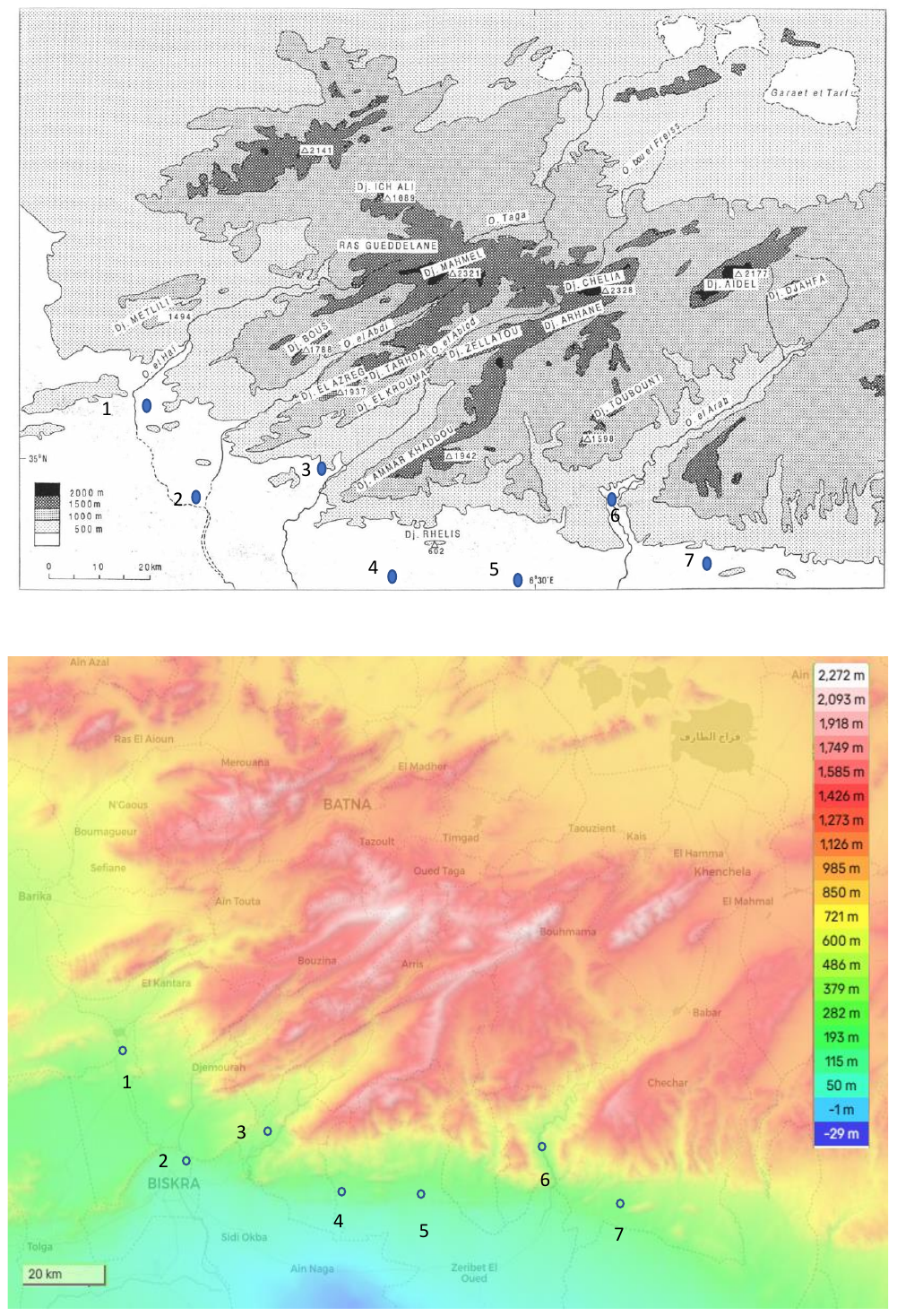}
	\caption{Carte topographie for aures mountains in Algeria  and the optimal position of telescopes for tracking the neutrinos tau }
\end{figure}
\newpage
After calculating the effective area of neutrinos  and the effective area of tau and by  taking into account the flow of ultra  high-energy neutrinos    passing through the surface of the Earth, we were able to find how many events are expected in  Cherenkov Telescope Array (CTA)  to be seen in  our site (Figure 12)compared with the work of Damiano Fiorillo et all[28].

\begin{figure}[h]
	\centering
	   \includegraphics[width=0.70\linewidth]{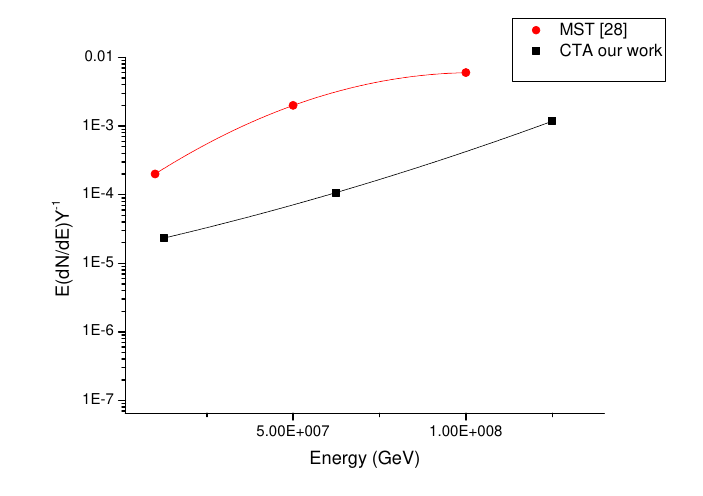}
	\caption{Differential number of events in one $cm^{2}$ per unit energy per unit time  expected at  CTA  in our cite compared with [28] }
\end{figure}
\newpage
\section{Conclusion}

This study explores the feasibility of capturing high-energy neutrinos by tracking Cherenkov photons emitted by the products of tau lepton reactions in the atmosphere. These tau leptons originate from neutrino interactions within mountain ranges. Through our investigation, we identified optimal characteristics for mountain ranges to achieve maximum effective area ratios, facilitating the capture of tau neutrinos around mountain perimeters of varying dimensions. We developed a Monte Carlo program capable of tracking the various stages of high-energy neutrino interactions. Our findings suggest that the most suitable mountain ranges should exceed dimensions of  40 x 40 $km^{2}$with heights exceeding 2 km. Utilizing simulations with the Corsika software, we determined optimal telescope locations for detecting Cherenkov photons around mountain ranges, ensuring comprehensive capture of all traces resulting from tau neutrino interactions. This involved tracking the flow of Cherenkov photons emitted by tau leptons at different altitudes and energies ranging from  $10^{7}-10^{8}$ GeV and at an angle of approximately  90 degrees. The results indicate that telescope heights should be  $ \leq$400 m , with a distance between two telescopes ranging from 12 to 17 km, and a distance between the telescope and mountain of about 10 to 18 km. In Algeria, we identified the Aures Mountains as the most suitable geographical location meeting conditions close to those outlined in Table 5. The specific locations for equipment placement are detailed in Figure 11 and Table 5. Finally, we estimated the expected number of events per unit energy per unit time in one  $cm^{2}$ per unit energy per unit time  expected at  CTA  in our cite  shown in figure 12.

%\end{multicols}
\end{document}